\documentclass[aps,prd,10pt,nofootinbib,twocolumn,eqsecnum,showpacs,showkeys,superscriptaddress,preprintnumbers,altaffilletter]{revtex4-1}
\usepackage{graphicx}
\usepackage{slashed}
\usepackage{amsmath}
\usepackage{multirow}
\usepackage{hyperref}
\usepackage{soul}
\usepackage{graphicx}
\usepackage{dcolumn}
\usepackage{amssymb}
\usepackage{amsfonts}
\usepackage{amsbsy}
\usepackage{color}
\usepackage{rotating}
\usepackage[english]{babel}
\usepackage{multirow}
\usepackage{float}

\usepackage{animate}

\newcommand{\be}{\begin{equation}}
\newcommand{\ee}{\end{equation}}
\newcommand{\bea}{\begin{eqnarray}}
\newcommand{\eea}{\end{eqnarray}}
\newcommand{\beal}{\begin{aligned}}
\newcommand{\eeal}{\end{aligned}}
\newcommand{\bi}{\begin{itemize}}
\newcommand{\ei}{\end{itemize}}

\hypersetup{
    colorlinks=true,
    linkcolor=blue,
    filecolor=magenta,
    urlcolor=cyan,
    citecolor=cyan
}

\begin{document}

\title{Weinberg's theorem, phantom crossing and screening}

\author{Philippe Brax}
\email{philippe.brax@ipht.fr}
\affiliation{Institut de Physique Th\'{e}orique, Universit\'{e} Paris-Saclay, CEA, CNRS, F-91191 Gif-sur-Yvette Cedex, France}

\date{\today}

\begin{abstract}
We consider models where the dilaton, seen as the pseudo-Goldstone boson of broken scale invariance, plays the role of dark energy. We revisit Weinberg's theorem and show that quantum corrections induced by the graviton lead to the screening of the dilaton locally. We also discuss the time evolution of the equation of state and find that phantom crossing is a natural feature of these models. The time variation of the equation of state and its deviation from $-1$ is limited by screening locally and can only be relaxed when the dilaton is allowed to have a mass of the order of the Hubble rate cosmologically, thus going beyond single-field screened dark-energy models. This obstruction extends to all single-field screened models of the chameleon-type where the large mass of the scalar on cosmological scales leads to a negligible variation of the equation of state at low redshift.  

\end{abstract}

\maketitle

\section{Introduction}
Dark energy is a late time probe of high energy physics \cite{Martin:2012bt,Polchinski:2006gy,Agmon:2022thq,McAllister:2023vgy}. As such, the recent results by the DESI experiment \cite{DESI:2025zgx} and the future measurements by large-scale surveys \cite{Amendola:2016saw} offer an unprecedented window on physics beyond the standard model of particle physics and cosmology \cite{Brandenberger:2025hof}. These experiments test in an indirect manner  energy scales corresponding to the UV completion of the standard model. This sensitivity to UV physics manifests itself primarily in the discrepancy between the expected value of the vacuum energy, which ought to be related to the UV (Ultra Violet) completion scale, and the minute value inferred cosmologically. This is the famous cosmological constant problem \cite{Weinberg:1988cp,Burgess:2004kd} usually interpreted as our theoretical inability to predict the observed dark energy amount.

Dark energy is also extremely sensitive to standard model physics. Indeed,  the quantum fluctuations from all massive particles,  including the ones from the standard model,   contribute to the vacuum energy\footnote{See appendix \ref{app:vac} for a calculation of the vacuum energy in the Decoupling Minimal Subtraction scheme.}. For instance the vacuum fluctuations due to the electron are much larger than the observed vacuum energy by around  thirty five orders of magnitude.  How do the UV part of the vacuum energy and the standard model contributions  compensate each other and  lead to the measured vacuum energy  is one of the startling aspects of the cosmological constant problem.

The low value of the vacuum energy can also be seen as a clue towards UV properties which would be difficult to access at the relatively low energies reached by particle accelerators\footnote{The Hierarchy Problem for the Higgs mass falls in the same category.}. Two possible consequences of this hint could be that the UV completion is either almost scale invariant or satisfies some degree of supersymmetry \cite{Burgess:2013ara,Burgess:2021obw}, and that these features are somehow preserved all the way down to the IR (Infra Red) regime of the Universe. This would guarantee that the vacuum energy could be either vanishing or hopefully small compared to the natural scales of the UV completion.

These ingredients are at the heart of one of the only firm results on dark energy gleaned over the last 40 years, i.e. Weinberg's no-go theorem stating that a fine-tuned value of the vacuum energy is inevitable provided that nature is described by a four-dimensional quantum field theory and that the vacuum is Poincar\'e invariant \cite{Weinberg:1988cp}. We will revisit the proof of this theorem as it gives hints of possible solutions to the cosmological constant problem, e.g. the vacuum may not be described by a static configuration in field space but may become dynamical \cite{Wetterich:1987fm,Ferreira:1997hj,Copeland:1997et,Barreiro:1999zs,Zlatev:1998tr,Steinhardt:1999nw,Wang:1999fa,Binetruy:1998rz,Brax:1999gp,Anchordoqui:2025fgz} leading  to the existence of dynamical dark energy. Alas, dynamical dark energy has bad press \cite{Efstathiou:2025tie}. Indeed, the cosmological dynamics in its recent accelerated phase would have to be dominated by a light field, often thought to be a scalar field, whose mass would have to be of the order of the present Hubble rate. This follows from requiring that its time variations are on  scales of the age of the Universe. Such a small mass is problematic as it suffers from a hierarchy problem due to quantum corrections from particles of much larger masses, such as the electron for instance. 

We will follow one of the proofs of Weinberg's theorem and link the smallness of the dark energy density to the spontaneous breaking of scale invariance. In this setting, one scalar field, i.e. the dilaton, emerges as the pseudo-Goldstone boson of broken scale invariance. In such models, quantum corrections are a blessing as they are responsible for the runaway potential which eventually leads to dark energy. We will study the effect of stabilising the dilaton by giving it a vacuum expectation value (vev) and  a mass. This combines with the exponential runaway behaviour to form an Albrecht-Skordis\footnote{Also used in the Yoga models \cite{Burgess:2021obw} } model of dark energy \cite{Albrecht:1999rm}. In these models, the dark energy scale and the effective dilaton mass in the Einstein frame are determined by the exponential rescaling from the Jordan to the Einstein frame. They can be made small by switching on appropriately the UV dilaton potential scales in the Jordan frame, which are then scaled down in the Einstein frame.  When these scales are small, the dilaton mediates a fifth force and as such would be extremely constrained in the solar system \cite{Bertotti:2003rm,Williams:2012nc}. 

Another UV sensitivity, which is   often overlooked,  arises from graviton loops. Indeed dark energy scalars with a very small mass interact with gravity and gravitation itself affects massive fermions. This implies that an induced coupling between dark energy and matter appears gravitationally. This happens at the one loop level and leads to a logarithmic divergent coupling. Of course, this divergence only indicates that the effective coupling between dark energy and matter is determined at the UV completion scale and depends on the matching with high energy physics. When applied to the dilaton model, we find that a quadratic coupling of the dilaton to matter is generated. This is the coupling normally postulaled for the environment-dependent dilaton \cite{Brax:2010gi,Brax:2011ja,Brax:2022uyh}. It  leads to screening of fifth forces locally and therefore helps evading gravitational tests in the solar system for instance.  Here we do find that the dilaton becomes very massive in dense environments. This feature remains true  on cosmological scales where the matter-dependent term in the effective mass leads to larger values than the Hubble rate. In fact, in these coupled models, the mass of the dilaton contains two parts. The mass in vacuum can be small and is affected by quantum corrections in a way related to the corrections to the vacuum energy, i.e. implying that only one tuning takes place \footnote{At least at the one loop order in the particle masses.}, whilst the density- dependent part is large when screening applies \cite{Brax:2012gr}. 
This has important consequences for the cosmological dynamics of the dilaton. Indeed, 
another effect derived  from the induced coupling of the dilaton to matter is the possibility of phantom crossing. This takes place when the equation of state of dark energy becomes smaller than $-1$. This can happen for coupled models \cite{Das:2005yj,Brax:2011qs,Martin:2005bp,Andriot:2025los,Wolf:2025jed,Khoury:2025txd} and we find that this is the case for the dilaton. Unfortunately, the large mass due to screening prevents a large variation of the equation of state with time. A substantial variation can only be achieved when the dark energy scalars are light cosmologically together with an unscreened  coupling to matter on cosmological scales. Generically, this is in conflict with standard screening in the single-field case \cite{Brax:2012gr,Brax:2021wcv} and seems to point out that phantom crossing could emerge more naturally in multi-field screened models \cite{Smith:2025grk}.

The paper is arranged as follows. In a first part we revisit the link between broken scale invariance, Weinberg's theorem and the dilaton. In section III, we consider how the dilaton can be related to dynamical dark energy and phantom crossing. Then in section IV, we discuss the gravitationally induced matter couplings and how they affect the dilaton's dynamics. We conclude with a discussion in section V. We have also added several appendices about more technical points. In particular, we discuss a new and simple parameterisation of the equation of state. 

\section{Weinberg's theorem revisited} 
We consider a theory with $N+1$ scalar fields coupled to gravity. We will assume that the dynamics of the scalars are governed by a dimension-four action with no intrinsic scale.   In particular, the scalar potential $V(\phi^I)$ is such that $V(\lambda \phi^I)= \lambda^4 V(\phi^I)$. This global scale invariance is taken as an accidental symmetry of the model at energies below the cut-off scale $\Lambda_{\rm UV}$ and is broken by the vacua of the potential leading to the existence of a Goldstone mode, i.e. the dilaton. Notice that the effective cosmological constant coming from integrating out modes of energy larger than $\Lambda_{\rm UV}$ vanishes. The action reads explicitly
\be 
S= \int d^4 x \sqrt{-g}(\frac{h(\phi^I)}{2}R - \frac{\sigma_{IJ}}{2} \partial_\mu \phi^I \partial^\mu \phi^J -V(\phi^I))
\ee
where the $\sigma$-model metric is chosen to be $\sigma_{IJ}=\delta_{IJ}$ and the function $h(\phi^J)$ is such that $h(\lambda \phi^I)=\lambda^2 h(\phi^I)$. The potential depends on a set of dimensionless couplings $g_\alpha$ corresponding to  the possible quartic interactions between the scalars. The scalars are coupled to fermions representing matter fields which eventually would comprise for instance the electrons or dark matter
\be 
S_\psi= -\int \sqrt{-g}( i k_{ab} \bar \psi^a \slashed{D} \psi^b + \lambda_{Iab} \phi^I \bar \psi^a \bar \psi^b)
\ee
where $D_\mu$ is the covariant derivative acting on the spinors and $k_{ab}$ is a $\sigma$-model metric for the fermions taken to be the identity matrix here for simplicity.  The Yukawa couplings $\lambda_{Iab}$ imply that the fermions become massive when the scalars condense. We do not introduce gauge degrees of freedom here. They could be added to make the setting more realistic. 

Let us first consider the vacua of this theory characterised by Poincar\'e invariant vacuum expectation values (vev's) for the scalars $\bar \phi^I$. Because of scale invariance in the scalar sector, we can always fix the value of one of the fields to the only scale in the problem, i.e. the Planck mass as determined by Newton's constant in the gravitational interactions at low energy, and consider the dynamics for the remaining $N$ fields, i.e.
\be 
\phi^i=\phi^0 z^i, \ i=1\dots N
\ee
where $z^i$ is dimensionless. As a result,  the potential reads
\be
V(\phi^0,\phi^0 z^i)= (\phi^0)^4 \tilde V(z^i)
\ee
where we have defined the reduced potential for the rescaled fields $\tilde V(z^i)= V(1,z^i)$. The vacua are such that $\partial_{\phi^0} V= \partial_{\phi^i} V=0$ implying that
\be 
\partial_{z^i} \tilde V(z^i)=0, \ \ \tilde V(z^i)=0
\ee
where $\phi^0$ is not determined and can be chosen to be $\phi^0=m_{\rm Pl}$\footnote{ This breaks scale invariance. With this choice, the Einstein-Hilbert term becomes normalised if the function $h(\phi^I)$ is normalised such that $h(1,z^i)=1$. This corresponds to fixing $\phi^0$ in order to normalise Newton's constant at its experimental value.}. As can be easily seen, this is an over-constrained system of $N+1$ equations for $N$ fields implying that vacua exist only if the coupling constants satisfy a constraint 
\be 
f(g_\alpha)=0.
\ee
This is the first part of Weinberg's theorem: one cannot expect to have a vanishing vacuum energy without some degree of fine-tuning. 
This extra constraint between the couplings can be naturally realised, i.e. without unseemly tuning,  if a symmetry is imposed on the model. This is the case of global supersymmetry \cite{Wess:1992cp} which relates the couplings and guarantees that vacua have vanishing energy\footnote{ Unless supersymmetry is spontaneously broken and then the vacuum energy does not vanish anymore.}. In these models, the vacuum energy is automatically zero corresponding to the scale invariance of the scalar sector.

When vacua exist, the fermions acquire a mass defined by the mass matrix
\be
m_{ab}= m_{\rm Pl} (\lambda_{0ab}+\lambda_{iab}\bar z^i).
\ee
Similarly the scalars pick up a mass. This is important as quantum corrections modify  the potential after condensation and give a correction to the potential energy, see appendix \ref{app:vac}, 
\be 
\delta V= \frac{1}{64\pi^2} {\rm Str} (M_0^4\ln (\frac{\Lambda_{\rm UV}^2}{M_0^2} )
\ee
depending on the supertrace, i.e. the sum over the bosons minus the fermions, of the quartic power of the mass matrix $M_0$. Hence quantum mechanically, the cosmological constant does not vanish anymore. In fact, one must also take into account the energy deposited in the vacuum by phase transitions implying that the vacuum energy becomes
\be 
V_0= \frac{1}{64\pi^2} {\rm Str} (M_0^4\ln (\frac{\Lambda_{\rm UV}^2}{M_0^2} )+ V_{\rm PT}
\ee
where $V_{\rm PT}$ is associated to the phase transitions at energies below the cut-off scale.

So far we have missed a crucial ingredient in our analysis. Indeed we have fixed $\phi^0$ at the Planck scale but this is not dictated by the dynamics of the model. We could have chosen any other value. This is a reflection of the fact that there is a Goldstone mode in the model corresponding to changing the value of $\phi^0$ with no energy cost at all, i.e. this Goldstone mode is associated to a flat direction of the potential.

This can be extracted by considering a rescaling of the metric $g_{\mu\nu}$ serving as a proxy for rescalings of space-time distances. Let us then change variables and define \cite{Brax:2014baa}
\be
\phi^I= \rho y^I,\ \ g_{\mu\nu}= \frac{1}{\rho^2}g^J_{\mu\nu}.
\ee
The resulting action becomes\footnote{We use the identity 
\be 
R= \rho^2 ( R_J+ 6g_J^{\mu\nu} D_\mu^J D_\nu^J \ln \rho -6 g^{\mu\nu}_J \partial_\mu \ln \rho \partial_\nu \ln \rho)
\ee
and $\sqrt{-g}= \rho^{-4} \sqrt{-g_J}$.}
\bea
 S= \int d^4 x \sqrt{-g_J}&&(\frac{m^2_{\rm Pl}}{2}\rho^{-2}R_J- 3m_{\rm Pl}^2(\partial \ln \rho)^2 \nonumber \\ && -\frac{\sigma_{IJ}}{2\rho^2}\partial_{\mu}(\rho y^I)\partial^\mu(\rho y^J) -V(y^I))\nonumber \\
\eea
where we have normalised $h(y^0,y^i)=m_{\rm Pl}^2$ and 
vacua are still obtained as  $y^i= y^0 z^i$ with $\tilde V(\bar z^i)=0$. When the scalars have condensed, the effective action for the dilaton $\rho$ reads
\be
 S= \int d^4 x \sqrt{-g_J}(\frac{m^2_{\rm Pl}}{2}\rho^{-2}R_J- \frac{m^2_{\rm Pl}}{2}\gamma^{-2} (\partial \ln \rho)^2).
\ee
where  we have defined 
\be 
\gamma = (6+ x^2(1+\sigma_{ij}\bar z^i \bar z^j))^{-1/2}
\ee
and $x= \frac{y^0}{m_{\rm Pl}}$.
The sum is over $i=1\dots N$.
Normalising the dilaton as 
\be 
\rho= e^{-\gamma \frac{\phi}{m_{\rm Pl}}}
\ee
we find the effective action for the Goldstone mode
\be
 S= \int d^4 x \sqrt{-g_J}(\frac{m^2_{\rm Pl}}{2}e^{2\gamma \frac{\phi}{m_{\rm Pl}}}R_J- \frac{1}{2} (\partial \phi)^2).
\ee
As expected, the dilaton $\phi$ has no potential corresponding to a flat direction. On the other hand, the presence of the explicit breaking of scale invariance by the Planck scale implies that the Goldstone transformation
\be 
\phi\to \phi+c
\ee
where $c$ is a constant and corresponding to a change of scale is broken in the gravitational sector. In fact, we have obtained that at low energy below the scale at which all the scalars $z_i$ condense, the theory for the dilaton is a scalar-tensor theory. The coupling to fermions is obtained by substitution and we obtain
\be 
S_\psi= -\int \sqrt{-g_J}( i k_{ab} \bar {\tilde \psi}^a \slashed{D}_J \tilde\psi^b + y^0(\lambda_{iab} z^i + \lambda_{0ab})\bar \psi^a \psi^b)
\ee
where $\psi^a= \rho^{3/2} \tilde \psi^a$ and the covariant derivative is calculated with $g^J_{\mu\nu}$. As can be seen, the dilaton decouples from matter in this frame \cite{Brax:2014baa,Ferreira:2016kxi}. The complete theory can be identified with a tensor-scalar theory for a massless field, the dilaton $\phi$, coupled to matter via the Jordan metric related to the Einstein one as
\be 
g^J_{\mu\nu}=e^{-2\gamma \frac{\phi}{m_{\rm Pl}}} g^E_{\mu\nu}.
\ee
In the Einstein frame, the action reads
\be
 S= \int d^4 x \sqrt{-g_E}(\frac{m^2_{\rm Pl}}{2}R_E- \frac{1}{2}(1+6\gamma^2 ) (\partial \phi)^2).
\ee
implying that the normalised field in the Einstein frame is
\be 
\varphi= (1+6\gamma^2)^{1/2}\phi.
\ee
The two metrics are related by 
\be 
g^J_{\mu\nu}=e^{-2\beta \frac{\varphi}{m_{\rm Pl}}} g^E_{\mu\nu}
\ee
where
\be 
\beta= \frac{\gamma}{(1+6\gamma^2)^{1/2}}
\ee
which is small for small $\gamma$. When $\gamma$ is large corresponding to a scalar with no kinetic terms in the Jordan frame, we retrieve the famous $\beta= 1/\sqrt 6$ common to $f(R)$ theories \cite{Brax:2023nvx,Sotiriou:2008rp} and massive gravity \cite{deRham:2014zqa}.

The fermionic action becomes
\be 
S_\psi= -\int \sqrt{-g_E}( i k_{ab} \bar {\psi}^a \slashed{D}_E \psi^b + y^0e^{-\beta \frac{\varphi}{m_{\rm Pl}}}(\lambda_{iab} z^i + \lambda_{0ab}) \bar \psi^a \psi^b).
\ee
The fermionic mass matrix is
\be 
m_{ab}^F= y^0e^{-\beta \frac{\varphi}{m_{\rm Pl}}}(\lambda_{iab} z^i  + \lambda_{0ab})
\ee
Notice that this matrix scales as $e^{-\beta \frac{\varphi}{m_{\rm Pl}}}$. This is also the case for the scalar masses after condensation. As a result, the quantum corrections are now of the type 
\be 
\delta V= \frac{1}{64\pi^2} {\rm Str} (M_0^4\ln \frac{\Lambda_{\rm UV}^2}{M_0^2})e^{-4\beta \frac{\varphi}{m_{\rm Pl}}}
\ee
where $M_0$ is the mass matrix without the dependence on the dilaton. Taking into account the phase transition taking place in the Jordan frame, we have for the potential
\be
V(\varphi)= V_0 e^{-4\beta \frac{\varphi}{m_{\rm Pl}}}
\ee
where we assume that $V_0>0$.
Summarising, after quantum corrections, the dilaton has an action in the Einstein frame 
\be
 S= \int d^4 x \sqrt{-g_E}(\frac{m^2_{\rm Pl}}{2}R_E- \frac{1}{2} (\partial \varphi)^2  -V_0 e^{-4\beta \frac{\varphi}{m_{\rm Pl}}}).
\ee
corresponding to a runaway dilaton model \cite{Gasperini:2001pc} coupled to matter. In particular,  the fermions have a dilaton-dependent mass matrix.

The vacuum of the theory is now obtained for $\varphi \to \infty$ where the potential vanishes. The quantum corrections have lifted the flat direction of the dilaton and   all the particles have vanishing masses in the Poincar\' e invariant vacuum of the theory. This is obviously unphysical. This is the second part of Weinberg's theorem, i.e. the vanishing of the vacuum energy is only possible  after quantum corrections for an infinite value of the dilaton where all the particle masses vanish. Hence there is no Poincar\'e invariant vacuum with vanishing  energy and non-vanishing particle masses.

Of course, one can circumvent these negative results by breaking one of the assumptions of the theorem, i.e. one can look for dynamical configurations depending on time and therefore not Poincar\'e invariant. In this case the mass matrix of the fermion becomes field dependent and time-dependent when the dilaton becomes dynamical. This is what happens as the quantum corrections have lifted the dilatonic flat direction. The dilaton can roll down along its exponential potential and become a runaway dilaton. Eventually the true Poincar\'e invariant vacuum is asymptotically reached in a dynamical way.   This leads naturally to the existence of dynamical dark energy tied up to the spontaneous breaking of scale invariance, e.g fermions have masses. 

\section{Dynamical dark energy}

\subsection{Einstein versus Jordan}

The potential for the dilaton generated by the quantum corrections is of the runaway type and therefore perfect for dark energy. The main difference with a model of exponential quintessence is that the dilaton couples to all matter species, e.g. dark matter and electrons. This differs from coupled dark energy models \cite{Amendola:1999er} where the dark energy field is assumed to couple only to dark matter, this is not the case here. For a slowly rolling dilaton, particles acquire a time-dependent mass matrix. This has important phenomenological consequences. 

In the presence of cosmological matter of conserved density $\rho_m$, the matter density appearing in the Friedmann  and the Poisson equations is the Einstein frame density
\be 
\rho_m^E= e^{-\beta \frac{\varphi-\varphi_c}{m_{\rm Pl}}} \rho_m
\ee
where the cosmologically conserved matter density $\rho_m$ satisfies
\be 
\dot \rho_m + 3 H_E \rho_m=0.
\ee
This conserved density is identified with the Einstein density for a value of the field $\varphi=\varphi_c$. When the field $\varphi$ has a monotonic time evolution, this defines a unique time $t_c$ where 
$\rho_m^E(t_c)= \rho_m(t_c).$ Notice that the calibration density is a choice which eventually can be considered as a free parameter in the theory or when comparing with data should be taken as the density at the redshift where the identification with $\Lambda$-CDM is carried out.  This allows one to identify
\be 
\rho_m(t_c) = \sum_i m_{\psi}^i n_i(t_c)
\ee
where $n_i$ is the conserved density of the non-relativistic fermions comprising both dark matter and baryons. The mass $m_\psi^i$ is the physical fermion mass identified as the one measured in the laboratory at the present time $t_0$
\be 
m_\psi^i= e^{-\beta \varphi_0 /m_{\rm Pl}}m_{\psi 0}^{i}
\ee
where $m_{\psi 0}^{i}$ is the fermion mass in the Jordan frame.
For instance the electron mass is the one measured in the  laboratory $m_e$ obtained from the Jordan mass by the rescaling $e^{-\beta \varphi_0 /m_{\rm Pl}}$ where $\varphi_0$ is the field value at present time.  This fixes the normalisation of the number density for each species which then evolves as
\be 
n_i(t)= (\frac{a_E(t_c)}{a_E(t)})^3 n_i(t_c).
\ee

The dilaton's dynamics are governed by the effective potential \cite{Khoury:2003rn}
\be 
V_{\rm eff}(\varphi)= V_0 e^{-4\beta \frac{\varphi}{m_{\rm Pl}}}+(e^{-\beta \frac{\varphi-\varphi_c}{m_{\rm Pl}}}-1) \rho_m
\ee
such that the Klein-Gordon equation in the Einstein frame becomes
\be 
\ddot \varphi+ 3H_E \dot \varphi + \frac{d V_{\rm eff}}{d\varphi}=0
\ee
where $H_E= d\ln a_E /dt_E$, $a_E$ is the scale factor and the Friedmann equation 
\be 
3 m^2_{\rm Pl}H_E^2= \rho_r+\rho_m + \rho_{\rm eff}
\label{Fried}
\ee
where $\rho_{\rm eff}= \frac{1}{2} \dot \varphi^2 + V_{\rm eff}(\varphi)$. Here dots denote time derivatives in the Einstein frame. Notice that the dependence on $\varphi$ in the Einstein density $\rho_m^E$ has been included in the effective potential whilst the conserved matter density has been made explicit in the Friedmann equation. 

The dynamics of this model with exponential potential and coupling are well-known \cite{Amendola:1999er}. We will extract some of the salient results of its solution. 
In the matter dominated era, the expansion of the Universe is slowed down with
\be 
a_E= a_{\rm eq}(\frac{t_E}{t_{\rm eq}})^\alpha, \ \varphi= \varphi_{\rm eq }+ \delta m_{\rm Pl}\ln \frac{t_E}{t_{\rm eq}}
\ee
starting from the time of matter-radiation equality.
We find that
\be 
\alpha= \frac{2}{3}(1- \frac{2}{3} \beta^2), \ \delta= \frac{4}{3}\beta
\ee
in the small $\beta$ limit. Dark energy starts dominating at a time $t_\star$ where $\rho_{\rm eff}(t_\star)= \rho_m (t_\star)$ after which
\be 
a_E= a_{\star}(\frac{t_E}{t_{\star}})^{\tilde\alpha}, \ \varphi= \varphi_{\star }+ \tilde \delta m_{\rm Pl}\ln \frac{t_E}{t_{\star}}
\ee
where
\be 
\tilde \alpha= \frac{1}{8\beta^2}, \ \tilde\delta= \frac{1}{2\beta}.
\ee
In this regime the equation of state of dark energy is constant 
\be 
\omega_{\rm DE}^E= -1+ \frac{16\beta^2}{3}
\ee
where $p_\varphi= \frac{1}{2} \dot\varphi^2 - V_0e^{-4\beta \varphi/m_{\rm Pl}}$ and $\omega_{\rm DE}^E= \frac{p_\varphi}{\rho_{\rm eff}}$. As expected for a quintessence model defined by a  potential, this is larger than $-1$.  Although we have obtained that the dilaton behaves like dark energy, there is no sign of phantom crossing. 

As the dilaton is not screened, cosmological observations depend on the physics in the Jordan frame. This is where atomic transitions are not altered by the presence of the dilaton. In this frame, the dark energy density and pressure receive corrections from their Einstein frame counterparts. 
The Jordan frame is defined by the metric
$g^J_{\mu\nu}= A^2(\varphi) g_{\mu\nu}^E$ where $A(\varphi)= e^{-\beta \varphi/m_{\rm PL}}$. In the Jordan frame, we have the Friedmann equation
\be 
3 m_{J, \rm Pl}^2H_J^2=\rho^J_{\rm DE}+ \rho_m^J +\rho_r^J
\ee
where $m_{J,\rm Pl}= m_{\rm Pl}/A(\varphi)$ is the time-dependent Planck mass in the Jordan frame.  The matter and radiation energy densities are $\rho_m^J$ and $\rho_r^J$ respectively. They are conserved in the Jordan frame and are related to the conserved Einstein energy and pressure as
\be 
\rho_m= A^3(\varphi) \rho_m^J, \rho_r^E= A^4(\varphi) \rho_r^J
\ee
where $a_J= A(\varphi) a_E$ and $dt_J= A(\varphi) dt_E$.
The dark energy density is given by \cite{Brax:2015lra}
\be 
\rho^J_{\rm DE}=\rho^J_\varphi + \frac{\epsilon_J (2- \epsilon_J)}{(1-\epsilon_J)^2}(\rho^J_m+\rho^J_r +\rho^J_\phi)
\ee
where we have introduced 
\be 
\epsilon_J= \frac{d\ln A}{d\ln a_J}. 
\ee
  The scalar pressure in the Jordan frame is given by
\bea 
p^J_{\rm DE}&=& p^J_\varphi+\frac{\epsilon_J}{1-\epsilon_J}(p^J_\varphi + p^J_r) \nonumber \\ && + \frac{1}{3(1-\epsilon_J)^2}(\epsilon_J -\frac{2}{1-\epsilon_J}\frac{d\epsilon_J}{d\ln a_J})\times (\rho^J_\varphi+ \rho^J_m + \rho^J_r)\nonumber \\
\eea
where $\rho_\varphi^E= A^4(\varphi) \rho_\varphi^J$ and $p_\varphi^E= A^4(\varphi) p_\varphi^J$.
The conservation of dark energy is satisfied
\be 
\frac{d \rho^J_{\rm DE}}{dt_J}= -3 H_J (p^J_{\rm DE}+ \rho^J_{\rm DE}).
\ee
The dark energy equation of state is then defined as 
\be 
\omega^J_{\rm DE}= \frac{p^J_{\rm DE}}{\rho^J_{\rm DE}}
\ee
which is not constrained to be greater than $-1$.
As a result, in coupled models there are two dark energy equations of states. We will focus on the   one adapted to the CMB and BAO, i.e. $\omega_{\rm DE}^J$. 

In the dark energy era where both matter and radiation are negligible, the equation of state in the Jordan frame becomes
\be 
\omega_{\rm DE}^J= (1-\epsilon_J)(\omega_{\rm DE}^E +\frac{1}{3(1-\epsilon_J)}(\epsilon_J -\frac{2}{1-\epsilon_J}\frac{d\epsilon_J}{d\ln a_J}) )
.
\ee
Phenomenologically $\epsilon_J$ must be small as particle masses cannot vary very much in the Einstein frame. In fact we have
\be 
\epsilon_J= \frac{d\ln m_F}{d \ln a_J}= \frac{1}{2}\frac{d  \ln G_N^F}{d \ln a_J}
\ee
which shows that a variation of Newton's constant in the Jordan frame contributes to the change of the equation of state.

In the dilaton case, we find that
\be
\epsilon_J= -4\beta^2
\ee
implying that
\be 
\omega_{\rm DE}^J= -1 + {\cal O}(\beta^4).
\ee
Hence in the Jordan frame the model behaves like $\Lambda$-CDM. This follows from the fact that in the Jordan frame the potential becomes a mere cosmological constant. Small corrections due to the dilaton kinetic energy are subdominant. 
As a result, if data eventually find that the equation of state is very close to $-1$, the dilaton model would then become a possible candidate as it reproduces the standard model of cosmology up to irrelevant deviations. Moreover, the fine-tuning of the cosmological constant here becomes a tuning of the time $t_\star$ which depends on the initial conditions for the field $\varphi$.

\subsection{Phantom crossing}

So far we have assumed that, in the matter sector,  scale invariance is an exact global invariance. Following the general philosophy that no global symmetry should exist in quantum gravity \cite{Agmon:2022thq}, we will assume that the dilaton is only a pseudo-Goldstone boson of the softly broken scale invariance in the matter sector. We take the action in the Jordan frame to be
\be
 S= \int d^4 x \sqrt{-g_J}(\frac{m^2_{\rm Pl}}{2}e^{2\gamma \frac{\phi}{m_{\rm Pl}}}R_J- \frac{1}{2} (\partial \phi)^2 - \frac{m_J^2}{2}(\phi-\phi_\star)^2).
\ee
where the dilaton is stabilised at a value $\phi_\star$ which depends on the UV physics. Its mass is defined as $m_J$.  In the Einstein frame, the scalar action becomes
\bea
 S= && \int d^4 x \sqrt{-g_E}(\frac{m^2_{\rm Pl}}{2}R_E- \frac{1}{2} (\partial \varphi)^2  \nonumber \\ && - (V_0 + \frac{m^2}{2}(\varphi-\varphi_\star)^2) e^{-4\beta \frac{\varphi}{m_{\rm Pl}}} - V_1 e^{-8\beta \frac{\varphi}{m_{\rm Pl}}}).\nonumber \\ 
\eea
where we have reinstated the corrections to the vacuum energy $V_0$ from the matter particles and also included the quantum corrections due to the dilaton itself which depend on
\be 
V_1= \frac{m^4}{64\pi^2}\ln \frac{\Lambda_{\rm UV}^2}{m^2}.
\ee
Notice that the scaling of the matter and dilaton corrections differ as the mass of the dilaton varies in $e^{-2\beta \frac{\varphi}{m_{\rm Pl}}}$ \footnote{The dilaton mass before quantum correction is in fact
\be 
m^2_\varphi= m^2(1 -\frac{4\beta (\varphi-\varphi_\star)}{m^2_{\rm Pl}}+ \frac{16\beta^2 (\varphi-\varphi_\star)^2}{m^2_{\rm Pl}}).
\ee 
As the field is always close to $\varphi_\star$, we neglect the linear and quadratic corrections compared to the constant in the bracket.}. 
We used the relation $\varphi= \frac{\gamma}{\beta} \phi$ and introduced
$m= \frac{\beta}{\gamma} m_J$. The potential becomes similar to the one introduced by Albrecht and Skordis \cite{Albrecht:1999rm} later used in the Yoga models \cite{Burgess:2021obw}.  We will comment on the scales involved and potential  tunings in the discussion at the end of the paper. 

The cosmology of this model depends on 
the effective potential for the dilaton 
\bea
V_{\rm eff}&&= (V_0 + \frac{m^2}{2}(\varphi-\varphi_\star)^2) e^{-4\beta \frac{\varphi}{m_{\rm Pl}}}+V_1 e^{-8\beta \frac{\varphi}{m_{\rm Pl}}}\nonumber \\ &&+(e^{-\beta \frac{\delta \varphi}{m_{\rm Pl}}}-1) \rho_m\nonumber \\
\eea
where $\delta \varphi= \varphi-\varphi_c$ due to the calibration of the matter density. 
The minimum of this potential is close to $\varphi_\star$ implying that dark energy depends eventually on the combination
\be 
V_{\rm DE}= V_0 e^{-4\beta \frac{\varphi_\star}{m_{\rm Pl}}}+V_1 e^{-8\beta \frac{\varphi_\star}{m_{\rm Pl}}}.
\ee
In these models, the tuning of the vacuum energy is reduced to the tuning of $V_{\rm DE}$. For  given $V_0$ and $V_1$ corresponding to a spectrum of particle masses below $\Lambda_{\rm UV}$, the value of $\varphi_\star$ must conspire to make $V_{\rm DE}$ small. Similarly the effective mass of the dilaton is obtained by expanding the effective potential to second order around  $\varphi_\star$ and is close to 
\be 
m^2_{\rm eff}= m^2_{\rm DE}+\frac{\beta^2 \rho_m}{m_{\rm Pl}^2}e^{-\beta\frac{\delta\varphi_\star}{m_{\rm Pl}}}
\ee
where
\be 
m^2_{\rm DE}=  e^{-4\beta \frac{\varphi_\star}{m_{\rm Pl}}}(m^2+ \frac{16 \beta^2 V_0}{m^2_{\rm Pl}})+ 64 e^{-8\beta\frac{\varphi_\star}{m_{\rm Pl}}}\frac{ \beta^2 V_1}{m^2_{\rm Pl}}.
\ee
Notice the small matter dependence when $\beta \lesssim 1$.
Dynamically the field evolves around $\varphi_\star$ and its behaviour depends on the
value of $m_{\rm eff}$ compared to the Hubble rate. As long as $m_{\rm eff}\ll H_E$ the mass term is negligible and the field $\varphi$ behaves like a dilaton with no stabilising potential as described in the previous section. After the transition occuring when  the mass becomes equal to the Hubble rate, i.e.   $H_E=m_{\rm eff}$ and  subsequently when $m_{\rm eff}\gg H_E$ the dilaton tracks the minimum of the potential close to $\varphi_\star$. 
In the following we will focus on cases where $m$ is small enough that the term in $V_1e^{-8\beta \frac{\varphi_\star}{m_{\rm Pl}}}$ becomes negligible in the effective potential. Similarly we consider that $m^2 \gg \frac{16\beta^2 V_0}{m^2_{\rm Pl}}$ implying that 
\be 
V_{\rm DE}\simeq V_0 e^{-4\beta \frac{\varphi_\star}{m_{\rm Pl}}}, \ m_{\rm DE}\simeq m e^{-2\beta \frac{\varphi_\star}{m_{\rm Pl}}}.
\ee
It is convenient to define $\theta= \varphi-\varphi_\star$. 
The minimum of the effective potential is given to leading order by
\be 
\theta (\rho_m)\simeq  \frac{\beta}{m^2_{\rm eff}m_{\rm Pl}}(4V_{\rm DE}+ \rho_m)
\ee 
where $\beta \theta(\rho_m)/m_{\rm Pl}\ll 1$ as long as $\beta \lesssim 1$ and $m_{\rm eff} \gtrsim H_E$.
It is important to notice that the density decreases with time and therefore the field value $\theta(\rho_m)$ decreases with time. As the coupling function itself $A(\varphi)$ decreases with $\varphi$ we find that $A(\varphi)$ increases with time. This will be at the origin of the phantom crossing \cite{Khoury:2025txd, Andriot:2025los}.

We can evaluate the effective potential at the minimum $V_{\rm eff}(\rho_m) \equiv V_{\rm eff}(\varphi(\rho_m))$
which reads
at linear order in $\beta \theta /m_{\rm Pl}$
\be 
V_{\rm eff}(\rho_m)=(1- \frac{4\beta}{m_{\rm Pl}}\theta(\rho_m) )V_{\rm DE} -\beta\frac{\delta \theta(\rho_m)}{m_{\rm Pl}}\rho_m
\ee
where $\delta\theta= \theta-\theta_c$.
In this regime, the kinetic energy is negligible. The potential itself reads
\be 
V(\rho_m)\equiv V(\varphi(\rho_m))=(1- \frac{4\beta}{m_{\rm Pl}}\theta(\rho_m) )V_{\rm DE}.
\ee
As a result, the equation of state of dark energy in the Einstein frame becomes at leading order
\be 
\omega_{\rm DE}^E= -\frac{1}{1 -\beta\frac{\delta \theta(\rho_m)}{m_{\rm Pl}}\frac{\rho_m}{V_{\rm DE}}}
\ee
or equivalently
\be 
\omega_{\rm DE}^E= -\frac{1}{1 -\frac{\beta^2(\rho_m -\rho_{mc})}{m_{\rm Pl}^2 m^2_{\rm eff}}\frac{\rho_m}{V_{\rm DE}}}
\ee
where $\rho_{mc}=\rho_m(t_c)$ is chosen as the matter density at the calibration time. We see immediately that the equation of state crosses the phantom divide at the calibration time. The equation of state now is larger than $-1$ and reads
\be 
\omega_{{\rm DE} 0}^E= -\frac{1}{1 -\frac{\beta^2(\rho_{m0} -\rho_{mc})}{m_{\rm Pl}^2 m^2_{\rm eff}}\frac{\rho_{m0}}{V_{\rm DE}}}
\label{ww}
\ee
as $\rho_{m0}< \rho_{mc}$ and $m_{\rm eff}$ depends on $\rho_{m0}$ weakly when $\beta\lesssim 1$. In the past and before the calibration time we have $\omega_\phi <-1$. It is noteworthy that DESI as a BAO (Baryon Acoustic Oscillation) experiment calibrates its analysis around a redshift of $z\approx 0.5$. This leads here to a natural crossing of the phantom divide around this redshift as announced by the DESI team \cite{DESI:2025zgx,Herold:2025hkb,Gonzalez-Fuentes:2025lei}\footnote{For a recent critique of the DESI analysis and the crucial role of the matter fraction, see \cite{Lee:2025hjw}.}. 

In practice, the analysis and comparison to data in the models presented here is complex. Indeed the DESI experiment uses multiple tracers \cite{DESI:2025zgx} from a redshift of $z_{\rm eff}=0.295$ for bright galaxies to $z_{\rm eff}= 2.33$ for the Lyman-$\alpha$ clouds. In principle, when analysed separately each tracer is associated to a calibration redshift when identifying $z_c\simeq z_{\rm eff}$. Typically we expect that $z_c \lesssim z_{\rm eff}$ as BAO measurements are sensitive to the Hubble rate  and the angular distance which take into account the integrated effect of dark energy between a redshift of $z=0$ and $z=z_{\rm eff}$. As such,  the Friedmann equation and the Hubble rate (\ref{Fried})  are thus associated to each tracer. This would lead to a different effective equation of state (\ref{ww}) for each tracer although in practice the equation of state is not measurable for each tracer as there are more cosmological parameters to contend with than data points. One would then expect to see the phantom crossing to take place at $z_{\rm eff}$ for each tracer. When combining all the data together one could consider $z_c$ as a parameter to be fitted. Grossly, it should be an averaged value of the $z_{\rm eff}$'s for the different tracers. Performing this analysis is beyond what we present here and is left for future work. See appendix \ref{app:par} for more details. For a recent result on the equation using a more general parameterisation than the CPL Chevallier-Polarski-Linder)  choice \cite{Chevallier:2000qy}, see \cite{Lee:2025pzo}.

In the Jordan frame, the equation of state  is corrected by terms depending on $\epsilon_J$ which reads here
\be 
\epsilon_J= 3\frac{\beta^2 \rho_m}{m^2_{\rm Pl}m^2_{\rm eff}}
\label{eps}
\ee
where we have neglected the small dependence on the density of $m_{\rm eff}$.
The effective equation of state is now given by
\be 
\omega_{\rm DE}^J= (1-3\frac{\beta^2 \rho_m}{m^2_{\rm Pl}m^2_{\rm eff}})(\omega_{\rm DE}^E- 5\frac{\beta^2 \rho_m}{m^2_{\rm Pl}m^2_{\rm eff}})
\label{state}
\ee
displacing the time of phantom crossing from the calibration time $t_c$ to the solution of
\be 
\omega_{\rm DE}^E= -1 +2\frac{\beta^2 \rho_m}{m^2_{\rm Pl}m^2_{\rm eff}}
\ee
corresponding a time slightly larger than $t_c$.

The fact that the equation of state in the Einstein frame crosses the phantom divide at the calibration time is general when the dark energy field tracks the minimum of the effective potential
\be 
V_{\rm eff}(\varphi)= V(\varphi) + (\frac{A(\varphi)}{A_c}-1)\rho_m
\ee
given by the solution to the equation
\be 
V'(\varphi)= -\frac{\beta(\varphi)}{m_{\rm Pl}}\frac{A(\varphi)}{A_c}\rho_m
\ee
where we have defined $\beta (\phi)= m_{\rm Pl}\frac{d\ln A(\varphi)}{d\varphi}$. Here $A_c= A(\varphi_c)$ associated to the calibration time $t_c$.
Let us solve for the evolution of the field $\varphi$ around the calibration time $t_c$. Defining $\theta= \varphi-\varphi_c$ we have
\be 
\theta(\rho_m)= -\frac{\beta(\varphi_c)}{m_{\rm Pl}}\frac{(\rho_m-\rho_{mc})}{m^2_{\rm eff}(\varphi_c)}
\ee
where the effective mass of the dark energy field is given by
\be 
m^2_{\rm eff}(\varphi_c)= \frac{d^2 V_{\rm eff}(\varphi)}{d\varphi^2}\vert_{\varphi=\varphi_c}.
\ee
In the dilaton case we have $\beta(\varphi)=-\beta$. When the mass of the dark energy field is much larger than the Hubble rate, the equation of state becomes
\be 
\omega_{\rm DE}^E= -\frac{1}{1 +\beta(\varphi_c)\frac{\theta(\rho_m)}{m_{\rm Pl}}\frac{\rho_m}{V_{\rm DE}}}
\ee
where we have identified $V_{\rm DE}= V(\varphi_c)$.
This becomes explicitly 
\be 
\omega_{\rm DE}^E= -\frac{1}{1 -\frac{\beta^2(\varphi_c)(\rho_m -\rho_{mc})}{m_{\rm Pl}^2 m^2_{\rm eff}(\varphi_c)}\frac{\rho_m}{V_{\rm DE}}}
\label{para}
\ee
crossing the phantom divide at $t_c$ in the Einstein frame. The more accurate result in the Jordan frame depends on $\epsilon_J$ where the previous expressions (\ref{eps}) and (\ref{state}) are valid with $\beta \to -\beta(\varphi_c)$. 
 In the past of $t_c$ when $\rho_m \gtrsim \rho_{mc}$, the equation of state becomes in the matter era
\be 
\omega_{\rm DE}^E \simeq -\frac{1}{1 -\frac{3\beta^2(\varphi_c)H^2} {m^2_{\rm eff}(\varphi_c)}\frac{\rho_m}{V_{\rm DE}}}
\label{state}
\ee
If the coupling $\beta(\varphi_c)={\cal O}(1)$, the deviation of the equation of state from $-1$ can become significant when 
$\rho_m \gg V_{\rm DE}$ compensates the smallness of $H^2/m^2_{\rm eff}(\varphi_c)$.

This type of  crossing of the phantom divide is common to all theories with a tracking minimum of the effective potential. This is in particular the case for all the models satisfying the chameleon mechanism \cite{Khoury:2003rn,Khoury:2003aq,Brax:2012gr}. In this case the coupling $\beta(\varphi_c)$ can be of order one whilst the local effect of the scalar field are screened. This is not what happens for the dilaton as $\beta$ is a constant and must be small to satisfy the Cassini constraint \cite{Bertotti:2003rm} in the absence of screening in the solar system.

We will now turn to the screening issue and show that a field-dependent part for the coupling of the dilaton to matter is induced by quantum loops of gravitons. This screens the dilaton locally. 

\section{Gravitationally induced matter coupling}

\subsection{The massive scalar}

Let us digress for a while before coming back to the dilaton. 
We consider a massless field $\varphi$ decoupled from fermions $\psi$. Typically we have in mind a situation like coupled quintessence \cite{Amendola:1999er} where the dark energy field couples to dark matter but not to matter fields. We will see that this is not tenable as gravitation intrinsically couples the quintessence field to the fermion $\psi$.
To see this let us expand the mass term to second order around a Minkowski background
\be 
\int d^4x \sqrt{-g} \frac{m^2}{2}\varphi^2 \supset 
-\int d^4x m^2 (1-\frac{h^2}{m^2_{\rm Pl}}) \varphi_0 \phi
\ee
where we have $g_{\mu\nu}=\eta_{\mu\nu} + \frac{2h_{\mu\nu}}{m_{\rm Pl}}$ and $\varphi=\varphi_0 + \phi$ where $\varphi_0$ is a background field value. We have $h^\mu_\mu=0$ and $h^2= h_{\mu\nu}h^{\mu\nu}$. Similarly the fermion mass term reads
\be 
\int d^4x \sqrt{-g} m_\psi \bar\psi \psi \supset \int d^4x (1-\frac{h^2}{m^2_{\rm Pl}}) m_\psi \bar\psi \psi 
\ee
Now the graviton propagator reads
\be 
\Delta_{\mu\nu\rho\sigma}(p)= \frac{i}{2 p^2}(\eta_{\mu\rho}\eta_{\nu\sigma}+ \eta_{\mu\sigma}\eta_{\nu\rho}-\eta_{\mu\nu}\eta_{\rho\sigma})
\ee
where $p$ is the four-momentum. The quadratic couplings between the graviton, matter and the scalar induce an interaction term at one loop order where a graviton runs inside it
\be 
\delta{\cal L}_{\rm coupling}=-\int d^4 x \frac{\beta(\varphi_0)}{m_{\rm Pl}} m_\psi \phi \bar\psi \psi
\ee
where the one loop coupling is given by
\be 
\beta(\varphi_0)= \frac{m^2 \varphi_0}{m^3_{\rm Pl}} \int \slashed{d}^4 p \Delta_{\mu\nu}^{\rho\sigma}(p)\Delta^{\mu\nu}_{\rho\sigma}(p)=\frac{2m^2 \varphi_0}{ m^3_{\rm Pl}}I
\ee
where the logarithmically divergent integral is
\be 
I= \int \frac{\slashed{d}^4 p}{p^4}= \frac{1}{8\pi^2}\int \frac{dp}{p}
\ee
where we have used that the volume of the 3d unit sphere is $2\pi^2$.
This implies that the fermion couples to the scalar field with the function
\be 
\beta(\varphi_0)= \frac{d\ln A}{d\varphi}\vert_{\varphi=\varphi_0}\Rightarrow A(\varphi)= 1+ 
\frac{m^2\varphi^2}{8\pi^2 m^4_{\rm Pl}}J
\ee
at leading order where $J= \int \frac{dp}{p}$.  So we find that the scalar field couples quadratically to matter.
If the interaction potential is $V(\varphi)$ and not a mass term, the interaction becomes 
\be 
A(\varphi)= 1+ 
\frac{V(\varphi)}{4\pi^2 m^4_{\rm Pl}}J
\ee
Several remarks are in order. First of all the order of magnitude $V(\varphi)/m^4_{\rm Pl}$ is small as we consider that Minkowski space-time is a good background around which the metric can be expanded. Of course this is upset by the fact that $I$ is infinite. So the result is of the type $0\times \infty$ and can only be understood after renormalisation. Indeed, the logarithmically divergent integral only tells us that the coupling between matter and the scalar is sensitive to the UV physics of the model. We can regularise the integral by introducing a UV cut-off $\Lambda_{\rm UV}$ and an IR one $\mu$ implying that
$ 
I= \ln \frac{\Lambda_{\rm UV} }{\mu}.
$
Defining the regularised coupling as
\be 
A(\varphi)= 1 + c_2 \varphi^2
\ee
we have 
the Callan-Symanzik equation in terms of the sliding scale $\mu$ 
\be 
\frac{d c_2}{d\ln \mu}= -\frac{m^2}{8\pi^2 m^4_{\rm Pl}}.
\ee
As the coupling constant $c_2$ is dimensionful, the renormalisation condition becomes
\be 
c_2(\Lambda_{\rm UV})= \frac{c}{\Lambda_{\rm UV}^2}
\ee
where $c$ is a constant and we have used the cut-off scale as the typical scale at the matching point. This implies that 
\be 
c_2(\mu)= \frac{c}{\Lambda_{\rm UV}^2} + \frac{m^2}{8 \pi^2 m^4_{\rm Pl}} \ln \frac{\Lambda_{\rm UV}}{\mu}.
\label{c2}
\ee
As expected there are two contributions to the coupling constant $c_2(\mu)$. The first one is the boundary terms at $\mu= \Lambda_{\rm UV}$ corresponding to integrating out all the UV scales beyond the cut-off $\Lambda_{\rm UV}$. This depends on the short distance physics and must be considered as an input coming from high energy. The second term is the running of the coupling at low energy $\mu \le \Lambda_{\rm UV}$. 
Cosmologically we are interested in the deep IR when $\mu\sim H \ll \Lambda_{\rm UV}$. 

The constant $c$ cannot be determined from the low energy point of view and must be matched with the coupling of the scalar $\varphi$ to matter in the UV completion. The logarithmic running is extremely small as it is suppressed by the Planck scale. In conclusion, the divergence of the graviton loop simply taught us that the coupling between the scalar and matter is a UV-sensitive quantity determined at the cut-off scale above which the UV completion must be considered. Interestingly this type of quadratic coupling seems to be favoured by data as shown in \cite{Wolf:2025jed}.

This result implies that quantum effects give a coupling to matter in coupled quintessence too even if classically the scalar field only couples to dark matter. In the absence of information on the UV completion, an agnostic attitude imposes that $c$ must be chosen of order unity. We will see the physical consequences of this below.

\subsection{The environment-dependent dilaton}

We can now come back to the dilaton and calculate the coupling to matter induced by graviton loops. As there is a bare coupling due to the breaking of scale invariance  determined by $\beta$,  we will consider the gravitational correction by  writing
\be 
A(\varphi)=A_0(\varphi)(1+ \delta(\varphi))
\ee
where we have denoted by $A_0(\varphi)=e^{-\beta\frac{\varphi}{m_{\rm Pl}}}$. This implies that 
\be 
\delta(\varphi)= \frac{V(\varphi)}{4\pi^2 m^4_{\rm Pl}} J
\ee
leading to the same divergence as in the previous section. 
Each term of the potential will have to be renormalised in this expression.
Let us come back to the dilaton. In this case, we can write, as a function of the sliding scale, 
\be 
\delta (\varphi)= c_0(\mu)+ c_2(\mu)(\varphi-\varphi_\star)^2
\ee
where
\be 
c_0(\mu)= c_0 + \frac{V_0}{4\pi^2 m^4_{\rm Pl}}\ln \frac{\Lambda_{\rm UV}}{\mu}
\ee
where $c_0$ is determined at the matching scale $\Lambda_{\rm UV}$. The expression for $c_2(\mu)$ is simply (\ref{c2}) again.

We can now write the effective potential of the dilaton model taking into account the new coupling function $\delta$. It reads\footnote{We still neglect the quantum corrections due to the dilaton mass and only keep the ones coming from the matter fields.}
\bea 
V_{\rm eff}(\varphi)&&= (V_0+ \frac{m^2}{2}(\varphi-\varphi_\star)^2)) A^4_0(\varphi)\nonumber \\  && + \frac{A_0(\varphi)}{A_0(\varphi_c)}(1+\delta(\varphi)-\delta(\varphi_c)) \rho_m\nonumber \\
\eea
after using the calibration of the matter density.
We have considered here that all matter species receive the same gravitational coupling, i.e. all the $c$ constants are equal. 
This model behaves like an environment-dependent dilaton with a mass depending heavily on matter \cite{Brax:2011qs}.

This can be seen explicitly by
expanding the potential around $\varphi_\star$ in $\theta= \varphi-\varphi_\star$. We have up to second order in $\theta$
\be 
V_{\rm eff}(\varphi)= V_{\rm eff}-\beta \frac{\theta}{m_{\rm Pl}} (4V_{\rm DE}+\rho_m) + \frac{m^2_{\rm eff}}{2}\theta^2
\ee
where the  energy density is 
\be 
V_{\rm eff}=V_{\rm DE} -\rho_m (\frac{\beta \theta_c}{m_{\rm Pl}}+ (c_2(\mu)+\frac{\beta^2}{2m^2_{\rm Pl}}) \theta_c^2)
\ee
and the effective mass is given by
\be 
m^2_{\rm eff}= m^2_{\rm DE} + \frac{\rho_m}{m^2_{\rm Pl}} ( \beta^2 + 2c_2(\mu) m_{\rm Pl}^2).
\ee
The effective potential is minimised for
\be 
\theta(\rho_m)= \frac{\beta( 4V_{\rm DE}+\rho_m)}{m_{\rm Pl}m^2_{\rm eff}}
\ee
When the matter density is the one of ordinary matter with $\rho_m \gg V_{\rm DE}$ then we have 
\be 
\theta (\rho_m) \simeq \frac{\beta}{2 m_{\rm Pl} c_2(\mu)}\simeq  \frac{\beta}{2c}\frac{\Lambda^2_{\rm UV}}{m_{\rm Pl}}
\ee
which is such that $\beta \theta(\rho_m)/m_{\rm Pl}\ll 1$ as $\Lambda_{\rm UV} \ll m_{\rm Pl}$ for $c={\cal O}(1)$.
We can now calculate the effective coupling to matter
\be 
\beta_{\rm eff}=m_{\rm Pl }\frac{d\ln A(\varphi)}{d\varphi}= - \beta + 2c_2(\mu)m_{\rm Pl}\theta
\label{thetainf}
\ee
which in dense matter becomes
\be 
\beta_{\rm eff}=0.
\label{beteff}
\ee
This is the expected result from a dilaton depending on the environment: its coupling to matter vanishes in dense matter. Hence we find that the coupling of the dilaton in dense matter is screened by graviton loops and that screening comes from the gravitational correction to the coupling function $A(\phi)$.

For finite size objects of large densities such as the earth, the coupling $\beta_{\rm eff}$ vanishes inside the object and varies over a thin-shell before converging to the bare value $-\beta$ outside the object. Screening is therefore not perfect and the effective coupling of such an object is given by the ratio \cite{Khoury:2003aq,Khoury:2003rn,Brax:2012gr,Brax:2021wcv}
\be 
\beta_{\rm eff}= \frac{\vert \theta_{\rm in}-\theta_{\rm out}\vert}{2m_{\rm Pl}\Psi}
\ee
where $\Psi$ is the Newtonian potential at the surface of the object. In a gedanken experiment, taking $\rho_m\to 0$ outside the objects, we have $\theta_{\rm out}= \frac{4 V_{\rm DE}}{m_{\rm Pl} m^2_{\rm DE}}$ which is negligible compared to $\theta_{\rm in}$ for large values of the cut-off scale $\Lambda_{\rm UV}$.
This implies that the screening factor is given by 
\be 
\frac{\beta_{\rm eff}}{\beta}=\frac{1}{4c \Psi}\frac{\Lambda^2_{\rm UV}}{m^2_{\rm Pl}}.
\ee
Screening necessitates that this ratio should be less than unity.

In fact the Cassini experiment requires that for the Sun with $\Psi_\odot \sim 10^{-6}$ we have $\beta_{\rm eff}^2 \le 2\times 10^{-5}$. Taking $c=1$ we have\footnote{A thorough investigation of the experimental constraints on screened dilatons can be found in \cite{Brax:2022uyh}.}
\be 
\Lambda_{\rm UV}\lesssim \frac{10^{-5}}{\sqrt{\beta}}m_{\rm Pl}.
\label{bound}
\ee
This implies that a hierarchy must exist between the cut-off scale $\Lambda_{\rm UV}$, the typical scale in the Lagrangian and the vevs of the scalar fields $y^i$ taken to be at the Planck scale\footnote{ Such a discrepancy is also present in axion models where the cut-off scale is typically smaller than the vev $f$ of the $U(1)$ breaking field.}.
Notice that $\Lambda_{\rm UV}$ has to be small enough to allow for a large mass of the dilaton in matter
\be 
m^2_{\rm eff}\simeq \frac{2c}{\Lambda_{\rm UV}^2}\rho_m.
\ee
Screening guarantees that for an object of size $R$ we have $m_{\rm eff} R\gtrsim  1$ as $m_{\rm eff}^2 R^2= 3 \frac{\beta}{\beta_{\rm eff}}\gtrsim 1$, which is at the origin of the chameleon screening for the dilaton in finite-size objects.

We can now express the equation of state of the screened dilaton in the Einstein frame. We focus on physical situations where the tracers of the matter densities are either stars in galaxies or molecular clouds where the density is larger than the dark energy density $V_{\rm DE}$. As the dilaton is fixed at a constant value by screening in such dense objects as soon as their density is much larger than the cosmological one, the Jordan and Einstein frames coincide up to an irrelevant rescaling implying that $\omega_{\rm DE}^E$ can be considered as the observable equation of state. This should apply to the DESI experiment where different calibration redshifts should be taken for the various tracers of the dark matter density. 

At the calibration point we have
\be 
\beta_{\rm eff} (\varphi_c)= \beta(-1+ 2 c_2(\mu) \frac{4V_{\rm DE}+\rho_c}{m^2_{\rm DE}(\varphi_c)})
\ee
where the mass of the dilaton is given by
\be 
m^2_{\rm eff}(\varphi_c)\simeq \frac{2c}{\Lambda^2_{\rm UV}}\rho_c
\ee
as the bound on $\Lambda_{\rm UV}$ in (\ref{bound}) implies that the density dependent part of the mass dominates over $m_{\rm DE}$ which is taken to be of the order of the Hubble rate $H_0$. This implies that
\be 
\beta_{\rm eff} (\varphi_c)\simeq 4\beta \frac{V_{\rm DE}}{\rho_c}.
\ee
We find that $\beta_{\rm eff} (\varphi_c)$ is of order $10 \beta$ when $z_c$ is small. Even for $\beta={\cal O}(1)$ the red-shift variation of $\omega_{\rm DE}^E$ is negligible due to the large effective mass from the screening criterion. Hence in these models, even though the dilaton is dynamical and dark energy differs from a cosmological constant, the variation of the equation of state away from $-1$ is so slight that phantom crossing would be hardly noticeable. 

\section{Discussion and conclusion}

We have studied dilaton models seen as a pseudo-Goldstone boson of broken scale invariance. After dilation breaking, the dilaton acquires a potential determined effectively by two energy  scales $V_{\rm DE}$ and $m_{\rm DE}$ determining the amount of dark energy and its mass. Both are related to the UV parameters  by a simple scaling
\be 
V_{\rm DE}\simeq e^{-4\frac{\beta}{m_{\rm Pl}}\varphi_\star}V_0, \ m_{\rm DE}\simeq e^{-2\frac{\beta}{m_{\rm Pl}\varphi_\star}} m
\ee
where $m$ is determined in the UV whilst $V_0$ depends on the supertrace of the quartic power of the matter mass matrix and the energy deposited in the vacuum by phase transitions. The mass $m_{\rm DE}$ is the mass of the dilaton in true vacuum when $\rho_m=0$. The effective mass of the dilaton $m_{\rm eff}(\varphi_c)$ receives a matter-dependent contribution which turns out to be larger than $m_{\rm DE}$ when screening is satisfied. Similarly, the fermions of the standard model pick up a mass of the order
\be 
m_F \simeq e^{-\frac{\beta}{m_{\rm Pl}}\varphi_\star} m_{\rm Pl}
\ee
implying that the scaling factor must be 
\be 
e^{-\frac{\beta}{m_{\rm Pl}}\varphi_\star}\simeq 10^{-15}
\ee
to obtain fermion masses in the TeV range with the Yukawa couplings determining the masses of each particles  from roundabout  $0.1$ for the top quark to much lower values for the electron. This determines the value of the stabilised dilaton $\varphi_\star$ as a function of the coupling $\beta$. The dark energy scale can be reproduced  if UV physics determines that it is related to the fermion masses as
\be 
V_0 \sim m_F^4\sim (10^3 {\rm GeV})^4.
\ee
This is far below the expected scale where $V_0\simeq m_{\rm Pl}^4$ in a model where all the scales are determined by the Planck scale. As an example, such a reduction of the vacuum energy can be achieved provided \cite{Burgess:2021obw} 
\be 
V_0 \simeq e^{-4\frac{\beta}{m_{\rm Pl}}\varphi_\star} m_{\rm Pl}^4.
\ee
Similarly, a mass $m_{\rm DE}\simeq H_0$ can  be achieved when
\be
m\simeq  \frac{\sqrt{V_0}}{m_{\rm Pl}}\simeq e^{-2\frac{\beta}{m_{\rm Pl}}\varphi_\star} m_{\rm Pl}
\label{scale}
\ee
implying that
$
m_{\rm DE}\simeq  e^{-4\frac{\beta}{m_{\rm Pl}}\varphi_\star} m_{\rm Pl}.
$
As can be seen in these scalings, the dark energy scale depends on two ingredients. The first  is the exponential scaling $ e^{-\frac{\beta}{m_{\rm Pl}}\varphi_\star}$ and the second is the constant $V_0$. In this approach, the vacuum mass of the dilaton is determined by $m$  which is also be related to $V_0$. We see that these models are successful once $V_0$ is tuned against $\varphi_\star$ to determine $V_{\rm DE}$ and this implies that the mass of the dark energy field is no more tuned than the vacuum energy, i.e. there is essentially only one tuning of $V_0$ versus $\varphi_\star$.

These scalings can only be achieved by going beyond the setting we have considered. In this paper  $V_0$ and $m$ are UV parameters which cannot be calculated from the low energy point of view and must be inferred from the UV. In some models, they may be the result of a dynamical relaxation induced by UV fields responsible for the breaking of scale invariance \cite{Burgess:2021obw}.  
For dilaton  theories as described here and at late times corresponding to a redshift less than two where the large scale surveys are relevant, in the effective equation of state of dark energy the ratio $\rho_m/V_{\rm DE}$ is never large enough to compensate for the fact that $m_{\rm eff}(\varphi_c) \gg  H_0$ for screening to take place. 
In these models phantom crossing does happen but is hardly noticeable. 

The only possibility of reconciling $\beta(\varphi_c)={\cal O}(1)$ and screening on short scales is to disentangle screening from the acceleration of the Universe, e.g by employing two fields at least \cite{Smith:2025grk}. In these models, the dilaton would be responsible for the acceleration of the Universe with a mass at late times of the order of the Hubble rate. When its coupling to matter is of order unity cosmologically, then phantom crossing does occur around the calibration time and a significant variation of the equation of state away from $-1$ can take place at small redshift. Now this requires that screening should involve another field, e.g. an axion, whose coupling to the dilaton would induce screening  without affecting the cosmological behaviour of the dilaton. As the dilaton is then light close to the present time, it would oscillate with a period of the order of the age of the Universe \cite{Smith:2025grk,Kessler:2025kju}. Phantom crossing would happen further back in the past when the density of matter becomes large enough. In these models and depending on the calibration time, the present equation of state of dark energy would be larger than $-1$. If phantom crossing is confirmed by forthcoming experimental results and the equation of state differs from $-1$  then  the study of the time evolution of the equation of state in multi-field models will be certainly fruitful.

The confirmation that the equation of state of dark energy differs from $-1$ would  jeopardise a host of models based on chameleon screening such as $f(R)$ \cite{Brax:2012gr} and approaches using the effective theory of dark energy \cite{Gubitosi:2012hu} with a $\Lambda$-CDM background.  On the other hand, if  future surveys indicate that the equation of state is very close to $-1$ with a very small time drift, then models like the screened dilaton presented here become relevant as they combine in one sweep a link between the cosmic acceleration and quantum corrections, together with screening of the dilaton fifth force locally. 
\begin{figure}
\centering
\includegraphics[height=5cm,width=0.35\textwidth]{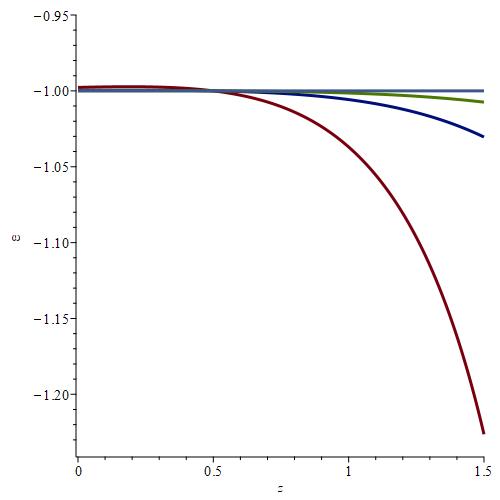}
\caption{The variation of the equation of state $\omega_{\rm DE}^E$ at low redshift for $\beta=0.1$ and four values of $\frac{m_{\rm eff}}{H_0}=(2,5,10,10^3)$. When the ratio is equal to $10^3$ the variation is negligible. It grows as the ratio decreases. Phantom crossing takes place at the calibration redshift $z_c=0.5$ here. }
\end{figure}

\appendix

\section{Vacuum energy}
\label{app:vac}

The vacuum energy is influenced by the quantum corrections due to massive particles. As a coupling of dimension four, the vacuum energy runs with the sliding scale $\mu$ from a value $\rho_{\rm UV}$ determined by the UV completion at energy scales $\mu \ge \Lambda_{\rm UV}$ to an IR value $\rho_{\rm IR}$ which determines one part of $V_0$. The other part in $V_0$ comes from the energy deposited by phase transitions at energies $\mu \le \Lambda_{\rm UV}$. The calculation of the contribution to the vacuum energy of a particle of mass $m$ can be obtained in the Decoupling Minimal Subtraction scheme \cite{Binetruy:1980xn,Burgess:2020tbq} where
\be 
\frac{d\rho_{\rm vac}}{d\ln\mu}=(-1)^{2s+1}\frac{m^4}{32\pi^2}\Theta(\mu-m)
\ee
where $\theta$ is the Heaviside distribution and $s$ the spin of the particle. In particular, this gives the intuitively pleasing result that below the threshold $\mu=m$ where particles of mass $m$ cannot be created, their quantum fluctuations are not efficient. As a result we find that at low energy $\mu \le m_{\rm min}$ where $m_{\rm min}$ is the lowest particle mass, e.g. this could be the neutrino mass, we have
\be 
\rho_{\rm IR}= \rho_{\rm UV}+ \frac{1}{32\pi^2} {\rm Str}(M^4_0\ln \frac{\Lambda_{\rm UV}}{M_0})
\ee
given by the supertrace over all the particles of masses $M_0$ lower that $\Lambda_{\rm UV}$. One can immediately see the origin of the cosmological constant problem. Indeed the UV contribution $\rho_{\rm UV}$ coming from integrating out the high energy modes must compensate terms at lower energies, such as the $m_e^4$ energy density from the electron fluctuations. Such an interplay between UV and IR physics is uncommon and corresponds to the absence of decoupling of UV effects at low energy. Moreover, conventional physics from the standard model of particles give large contributions whose cancellation is hard to explain. In the main text, we assume that $\rho_{\rm UV}=0$ as befitting certain supersymmetric string constructions and use the vacuum contributions from all the particle spectrum below $\Lambda_{\rm UV}$ to seed the potential for the dilaton.

\section{A new parameterisation of the equation of state}
\label{app:par}

In this appendix, we make explicit that a new parameterisation of the equation of state can be obtained by relaxing the assumptions in the main text. We propose to use (\ref{para}) and consider it valid for all values of the three parameters $(z_c, \beta(\phi_c), m^2_{\rm eff}(\phi_c))$. We simplify the notation and write it as
\be 
\omega_{\rm DE}^E= -\frac{1}{1 -\frac{\beta^2(\rho_m -\rho_{mc})}{m_{\rm Pl}^2 m^2_{\rm eff}}\frac{\rho_m}{V_{\rm DE}}}
\ee
which can be expressed as a function of the redshift in the Einstein frame denoted here by $z$
\be 
\omega_{\rm DE}^E=-\frac{1}{1-\frac{3\beta^2 \Omega_{m0}^2}{\Omega_{\Lambda 0}}\frac{H_0^2}{m^2_{\rm eff}}((z+1)^3-(z_c+1)^3)(1+z)^3}
\ee
which depends on the matter fraction $\Omega_{m0}$, the dark energy fraction $\Omega_{\Lambda 0}$, the coupling to matter $\beta$ and the ratio between the dark energy mass and the Hubble rate $m_{\rm eff}/H_0$. In fact, there are only two relevant parameters $z_c$ and $\beta^2 H_0^2/m^2_{\rm eff}$. Typically we expect $\beta\simeq 10^{-1}$ would give a deviation of structure growth at the percent level \cite{Brax:2022uyh}. Screening in the single field case would impose $m_{\rm eff}/H_0 \gtrsim 10^3$ \cite{Brax:2012gr}. Lower values necessitate to extend the model to the multi-field case. Here we consider the previous expression as a convenient fitting function whose behaviour captures some of the physics of both dynamical dark energy and screening. This parameterisation leads to a pole at finite redshift where the equation of state diverges \cite{Andriot:2025los}.  Finally the calibration redshift $z_c$ is an experimental input which would have to be inferred from  data. Examples of the variation of the equation of state can be seen in Fig. 1. 

Expanding the equation of state around $z=z_c$ to linear order we obtain
\be 
\omega^E_{\rm DE}(z)\simeq -1 - \frac{9\beta^2 \Omega_{m0}^2}{\Omega_{\Lambda 0}}\frac{H_0^2}{m^2_{\rm eff}}(1+z_c)^3(z-z_c).
\ee
This can be compared to the same expansion of the parameterisation $\omega= \omega_0 +\omega_a (1-a)$ which reads 

\be 
\omega (z) \simeq -1 +\frac{\omega_a}{(1+z_c)^2}(z-z_c)
\ee
where we have phantom crossing for
\be 
\omega_0= -1 + \frac{z_c}{1+z_c} \omega_a.
\ee
This allows us to identify
\be 
\frac{\beta^2 H_0^2}{m^2_{\rm eff}} \simeq - \frac{\Omega_{\Lambda 0}}{9\Omega_{m0}^2(1+z_c)^5} \omega_a
\ee
Taking $z_c \sim 0.37$ from Fig.2 of \cite{DESI:2025fii} and a variation of $\omega_a$ between $-1.6$ and $-0.8$ befitting the analysis in Fig.1 of \cite{DESI:2025fii}
corresponding to $\omega_a \simeq -3.66 (1+\omega_0)$ where $(1+z_c)/z_c=3.66$, we find that typically
\be 
0.047 \lesssim \frac{\beta^2 H_0^2}{m^2_{\rm eff}}\lesssim 0.093
\ee
for $\Omega_{\Lambda 0}\simeq 0.7$ and $\Omega_{ m0} \simeq 0.3$.
Models with $\beta ={\cal O}(0.1)$ and $m_{\rm eff} \sim H_0$ are within the right ball-park. 
We leave the challenge of finding screened models that satisfy these bounds for future work.

\section{Initial conditions}
We have taken the dilaton as a low energy field. One would like to have a theoretical understanding from inflation to dark energy. A possibility would be to use the dilaton as the inflaton \cite{Burgess:2022nbx}. This is a difficult route as the inflation and dark energy scales are so far apart. If one assumes that the dilaton couples to another sector of the theory where inflation takes place, then we can use the results 
of the main text and replace $\rho\to -T_{\rm inf}=12m_{\rm Pl}^2 H_{\rm inf}^2.$ This is so large compared to late-time cosmological scales that this pins down the dilaton close to $\varphi_\star$ at a distance given by (\ref{thetainf}). This provides the initial condition for the dilaton field. Then in the radiation era and as long as reheating is taken to be instantaneous, the field will not evolve as the Hubble friction dominates over the small effective potential. Moreover, the coupling of the dilaton to matter being effectively zero during the radiation era (\ref{beteff}), the dilaton does not feel kicks when species decouple \cite{Damour:1994zq}. In the matter era, the dilaton starts evolving under the influence of the matter term in its effective potential although never very far from $\varphi_\star$. Eventually, the dilaton reaches the late-time regime described in the main text. A precise description of the time evolution of the dilaton \cite{Albrecht:2001xt} including tracking solutions during the radiation era  is left for future work. 

\section{Kinetic couplings}
The graviton loops also couple the kinetic terms for the dilaton and matter. This implies that the Lagrangian receives a correction 
\be 
\delta {\cal L}\supset -\frac{(\partial \varphi)^2}{8\pi^2m_{\rm Pl^2}}m_\psi \bar\psi \psi J
\ee
depending on the logarithmically divergent integral $J$. After renormalisation this induces a coupling
\be 
\delta {\cal L}\supset -(\frac{c_{\rm kin}}{\Lambda_{\rm UV}^4 }+\frac{1}{4\pi^2m_{\rm Pl}^4}\ln (\frac{\Lambda_{\rm UV}}{\mu})) \frac{(\partial \varphi)^2}{2} m_\psi \bar\psi \psi 
\ee
In matter this leads to a wave function renormalisation 
\be 
\delta {\cal L}\supset -(\frac{c_{\rm kin}}{\Lambda_{\rm UV}^4 }+\frac{1}{4\pi^2m_{\rm Pl}^4}\ln (\frac{\Lambda_{\rm UV}}{\mu}))\rho_m \frac{(\partial \varphi)^2}{2} 
\ee
In all cases we consider situations where $\rho_m/\Lambda_{\rm UV}^4\ll 1$, so we can neglect this effect. 
\acknowledgments

I would like to thank D. Andriot, E. Armengaud,  P. Valageas,  P. Vanhove  and A. Smith for important remarks.  I am particularly grateful to A. de Mattia for insightful comments. I would like to thank C. Burgess for his 2025 les Houches lectures.  

\bibliography{ref}

\end{document}